\documentclass[conference]{IEEEtran}
\IEEEoverridecommandlockouts
\usepackage{cite}
\usepackage{amsmath,amssymb,amsfonts}
\usepackage{algorithmic}
\usepackage{graphicx}
\usepackage{textcomp}
\usepackage{xcolor}
\usepackage{booktabs}

\usepackage{multirow}
\usepackage{ulem}
\usepackage{hyperref}
\hypersetup{hidelinks,
	colorlinks=true,
	allcolors=black,
	pdfstartview=Fit,
	breaklinks=true
}
\usepackage[capitalize, nameinlink]{cleveref}
\DeclareMathOperator*{\argmax}{arg\,max}

\def\BibTeX{{\rm B\kern-.05em{\sc i\kern-.025em b}\kern-.08em
    T\kern-.1667em\lower.7ex\hbox{E}\kern-.125emX}}
\begin{document}

\title{A Wireless-Assisted Hierarchical Framework to Accommodate Mobile Energy Resources}

\author{\IEEEauthorblockN{Pudong Ge\IEEEauthorrefmark{1},
Cesare Caputo\IEEEauthorrefmark{2}, Michel-Alexandre Cardin\IEEEauthorrefmark{2}, Anna Korre\IEEEauthorrefmark{3}, and Fei Teng\IEEEauthorrefmark{1}}
\IEEEauthorblockA{
\IEEEauthorrefmark{1} Department of Electrical and Electronic Engineering, Imperial College London \\ \IEEEauthorrefmark{2} Dyson School of Design Engineering, Imperial College London \\
\IEEEauthorrefmark{3} Department of Earth Science and Engineering, Imperial College London \\
}}

\maketitle

\begin{abstract}
The societal decarbonisation fosters the installation of massive renewable inverter-based resources (IBRs) in replacing fossil fuel based traditional energy supply. The efficient and reliable operation of distributed IBRs requires advanced Information and Communication Technologies (ICT) , which may lead to a huge infrastructure investment and long construction time for remote communities. Therefore, to efficiently coordinate IBRs, we propose a low-cost hierarchical structure, especially for remote communities without existing strong ICT connections, that combines the advantages of centralised and distributed frameworks via advanced wireless communication technologies. More specifically, in each region covered by a single cellular network, dispatchable resources are controlled via a regional aggregated controller, and the corresponding regional information flow is enabled by a device-to-device (D2D) communication assisted wireless network. The wireless network can fully reuse the bandwidth to improve data flow efficiency, leading to a flexible information structure that can accommodate the plug-and-play operation of mobile IBRs. Simulation results demonstrate that the proposed wireless communication scheme significantly improves the utilization of existing bandwidth, and the dynamically allocated wireless system ensures the flexible operation of mobile IBRs.
\end{abstract}


\section{Introduction}
The worldwide consensus towards carbon neutrality has been prompting a revolutionary development in the energy industry \cite{creutzig2014catching,ge2021cyber}. One of the most significant revolutions is the rapid development of renewable energy resources, especially inverter based resources (IBRs) \cite{apostolopoulou2016interface}. An increasing number of IBRs brings huge uncertainties and fluctuations, which imposes severe operational challenges to maintain the system stability and balance. The existing control frameworks are not robust nor reliable to manage increasing penetration of renewable generation. Therefore, it is urgent to coordinate all dispatchable energy resources in power systems for a secure, reliable and resilient energy supply framework \cite{wang2020sustainable} under the concept of ``\textit{cyber-physical power system (CPPS)}''.

Nevertheless, in many developing countries or low-income regions, the existing information infrastructure cannot support the intelligent operation of IBRs, which may require a huge infrastructure investment and long construction time. Take Mongolia as an example, as the population migration process, a category of suburban living areas called ``ger districts'' surrounding its capital Ulannbataar accommodate a large population of the country \cite{park2019spatiotemporal}, which promotes the renewable development in the local areas to tackle energy supply challenge. However, the remote communities also suffer from lacking of advanced information infrastructure. In addition, many Mongolian citizens are still living in a nomadic lifestyle, which requires a mobile form of energy supply. Hence, it is of great significance to consider the flexibility and investment cost of supporting information systems while developing renewable resources.

The information structure determines interacting mode among increasing IBRs and existing control centers (e.g. SCADA system). In general, there are two main types of coordination frameworks, i.e., centralised and distributed.
However, while widely studied and implemented \cite{cheng2018centralize,tsikalakis2011centralized}, a centralised control framework has its own drawbacks \cite{ge2020extended}: 1) the inevitable costs on the implementation of complex centralised communication mode due to long-distance remote communication from control center to remote renewable generators; 2) long-time and indispensable delays due to all-to-one communication and high-dimension computing; 3) low robustness and resiliency suffering from single-point failure; 4) low plug-and-play capability.

To overcome the above limitations, sparse-communication-based distributed control has been investigated. In a distributed control framework, each dispatchable agent communicates with neighbouring agents and calculates control commands locally. By utilising a distributed framework, control objectives in the CPPS, e.g., frequency control \cite{zhao2015distributed}, voltage control \cite{ge2020resilient,ge2021event}, energy management \cite{zia2018microgrids}, etc., have been discussed comprehensively via both distributed feedback control \cite{ge2020extended,qu2019optimal} and distributed optimisation{\cite{molzahn2017survey}}. The distributed framework has been demonstrated to have better scalability and resiliency, which is promising in accommodating a growing number of IBRs.
In fact, these small-capacity but massive-volume IBRs will be integrated into the CPPS with an aggregation mode in remote communities. Under this circumstance, if we adopt a fully distributed framework, massive numbers of IBRs will increase the complexity of distributed network and will incur huge implementation cost of replacing current centralised framework. Hence, one of the most significant challenges of efficiently and cost-effectively managing IBRs lies in the framework design, i.e., exploring a low-cost and efficient communication solution that has capability of being compatible with legacy devices and of fostering the future algorithm development in the CPPS.

To efficiently regulate massive and increasing IBRs, we propose a low-cost and flexible hierarchical structure that combines both advantages of centralised control and distributed control via the advanced wireless communication system such as 5G \cite{wu2017overview}. More specifically, based on the advanced wireless communication technology, massive dispatchable resources are controlled via regional aggregated controllers (e.g. base station associated control centers), and aggregated controllers of each community are cooperatively operated in a distributed manner to reach the real-time optimality using distributed feedback control. Even in remote communities or rural areas, IBRs can provide necessary energy supply with a cellular network. In addition,  current EMS or SCADA can still function in non-real-time applications, i.e., periodically implementing economical dispatch and optimisation to enhance the CPPS operational economy, and monitoring and diagnosing faults, etc. To summarise, the main contributions of the proposed hierarchical framework are concluded as
\begin{enumerate}
    \item We propose a hierarchical framework where an aggregated controller installed with one base station regulates regional IBRs. This design has common interests with increasing constructions of wireless devices, such as base stations, and provides a low-cost solution that offers control flexibility via a dynamic wireless network.
    \item A wireless resource allocation including device-to-device (D2D) communication is comprehensively analysed in terms of supporting renewable IBRs coordination. More specifically, the proposed D2D-assisted wireless communication allows plug-and-play operations from the perspective of cyberspace, based on which mobile IBRs can be effortlessly accommodated.
\end{enumerate}

The rest of this paper is arranged as below: \cref{sec:2} gives an overview of the proposed hierarchical framework, while \cref{sec:3} details the wireless communication resource allocation algorithm that supports the plug-and-play operation of mobile IBRs. \cref{sec:4} designs a demonstrated case to evaluate the wireless resource allocation algorithm, and \cref{sec:5} concludes the paper.

\section{Wireless Assisted IBR-Dominated CPPS and Its Hierarchical Control Framework\label{sec:2}}

\begin{figure}[!htb]
\centering
\includegraphics[width=\columnwidth]{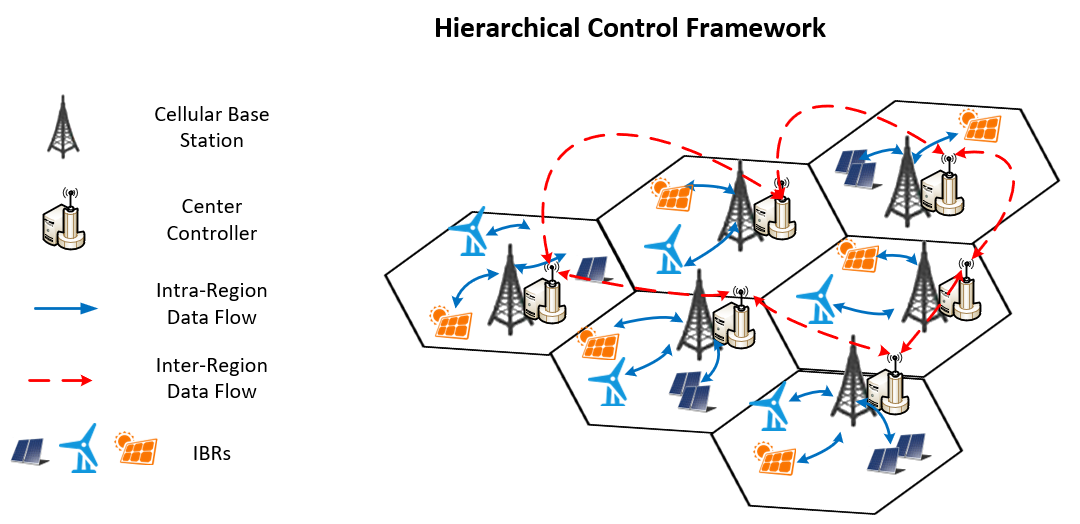}
\caption{A hierarchical framework assisted wireless communication to accommodate massive dispatchable IBRs.}
\label{fig:framework}
\end{figure}

In this section, a hierarchical control framework assisted by wireless technologies is elaborated comprehensively, as shown in \cref{fig:framework}. Each base station has its own coverage, IBRs within which are regulated as an aggregation, while co-existing base station and regional control center are responsible for information exchange and control algorithm implementation respectively.

\subsection{Intra-Community Aggregation}
\begin{figure}[!htb]
\centering
\includegraphics[width=0.7\columnwidth]{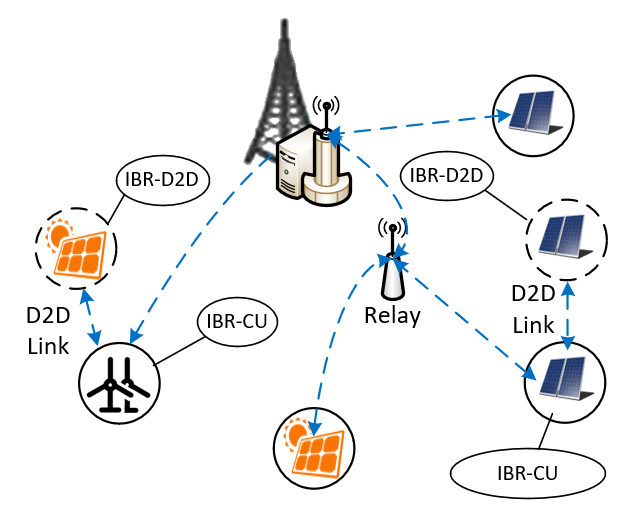}
\caption{Regional CPPS based on wireless and D2D technology}
\label{fig:region_D2D}
\end{figure}

Dispatchable IBRs are regulated in an aggregated way within a community. These dispatchable resources are equipped with a wireless module (antenna) enabling wireless information flows with base stations. The wireless communication topology can be time-varying by resource allocation algorithms \cite{wu2017overview} implemented inside base stations, which conducive to plug-and-play operation of mobile devices such as EVs. As a developing 5G technology, D2D communication \cite{asadi2014survey} can further increase the coverage of each base station. To specialize the generality of \cref{fig:framework}, the wireless communication is detailed in \cref{fig:region_D2D}. The D2D communication is one of the most promising way to accommodate the cyber links of IBRs without additional costs.

The dispatchable IBRs contain various elements, either mobile or non-mobile, including photovoltaic (PV), wind turbine, storage, EV, controlled load such as air conditioner, etc. The non-mobile IBRs (named IBR cellular users, i.e., IBR-CUs) are installed with a directed link with regional base station, while mobile IBRs (named IBR-D2Ds) are allowed to plug-and-play operation and linked to the cellular network via D2D communications.
The aggregation model participating in the CPPS regulation of IBRs has been widely investigated \cite{chen2021aggregated}, in terms of grid-forming (GFM) inverter and grid-following (GFL) inverter, and an aggregated dynamic model can be built through equivalent methods~\cite{roos2020aggregation,gholami2021stability}. Hence, if the communication solution, which is the focus of this paper, is well designed, then a regional aggregation can be easily conducted.

\subsection{Inter-Community Coordination}
As seen in \cref{fig:region_D2D}, a base station gathers regional operational information of IBRs in the coverage, and aggregates to form a regional dynamic model. For the inter-community coordination, the distributed communications among base stations (red dotted lines in \cref{fig:framework}) are enabled by optical fiber or wireless channel \cite{sharma2021review}, based on which many well-developed distributed algorithms \cite{ge2020extended,qu2019optimal,molzahn2017survey} can be implemented.
As such, the design of inter-community communication is not studied due to the limited space. Hence, hereafter, the regional wireless solution will be elaborated in details.
It should be noted that the proposed wireless-assisted hierarchical framework can be implemented in remote communities, where a cellular network is essential for daily communication. By using a cellular network for essential information flows and using D2D communications increasing bandwidth, the proposed wireless-assisted method will significantly reduce the investment cost on cyber-layer supporting devices, which is meaningful for low-income communities and developing countries.

\section{Regional Wireless Resource Allocation\label{sec:3}}
To detail the communication strategy of the proposed hierarchical framework based on wireless infrastructures, we first elaborate the regional information structure based on IBR-D2Ds, which is a promising way of enlarging the coverage and improving the performance of fully loaded cellular networks.

We will investigate D2D-assisted communications as in \cref{fig:region_D2D}, where $M$ IBR-D2Ds coexist with $N$ IBR-CUs. We use sets $\mathcal{D}=\{1,2,\dots,M\}$ and $\mathcal{C}=\{1,2,\dots,N\}$ denoting IBR-D2Ds and IBR-CUs. D2D communications are used to accommodate the newly developing renewable generators even if the exsting cellular network is already fully loaded . Hence, we assume $M<N$ which ensures that each IBR-D2D can find a reuse partner of IBR-CU. This assumption is reasonable because the capacity of the cellular network should be upgraded once D2D communications cannot satisfy the data rate requirement of renewable information flow. However, the capacity upgrade is out of scope and will be discussed in future works.
For the wireless spectrum allocation problem, we assume, $N$ active CUs occupy the $N$ orthogonal channels in the cell and there is no spare spectrum. We start the analysis from a single IBR-D2D. Only when the minimum SINR (Signal to Interference plus Noise Ratio) is met and the interference to IBR-CUs (i.e. non-D2D links) is below a threshold, a IBR-D2D can be successfully set up. As such, we call it a D2D candidate and the corresponding IBR-CU as a reuse partner. Hence, the overall optimization problem can be formulated as follows:
\begin{subequations}
    \begin{align}
        & \max_{\rho_{i,j},P_{i}^{c},P_{j}^{d}}\quad \sum_{i\in\mathcal{C}}\sum_{j\in\mathcal{D}}{\big[\log_2(1+\epsilon_{i}^{c})+\rho_{i,j}\log_2(1+\epsilon_{j}^{d})\big]} \tag {2} \label{eq:ra_obj}\\
        & \mathrm{subject\ to:} \notag \\
        & \qquad\epsilon_{i}^{c} = \frac{P_i^cg_{i,B}}{\sigma^{2}+\rho_{i,j}P_{j}^{d}h_{j,B}}\geq\epsilon_{i,min}^{c}, \forall\;i\in\mathcal{C} \label{eq:ra_c1}\\
        & \qquad\epsilon_{i}^{d} = \frac{P_i^dg_{j}}{\sigma^{2}+\rho_{i,j}P_{i}^{c}h_{i,j}}\geq\epsilon_{j,min}^{d}, \forall\;j\in\mathcal{D} \label{eq:ra_c2}\\
        & \qquad\sum_{j}{\rho_{i,j}}\leq1,\rho_{i,j}\in\{0,1\},\forall\;i\in\mathcal{C} \label{eq:ra_c3}\\
        & \qquad\sum_{i}{\rho_{i,j}}\leq1,\rho_{i,j}\in\{0,1\},\forall\;j\in\mathcal{D} \label{eq:ra_c4}\\
        & \qquad P_{i}^{c}\leq P_{\max}^{c},\forall\;i\in\mathcal{C} \label{eq:ra_c5}\\
        & \qquad P_{j}^{d}\leq P_{\max}^{d},\forall\;j\in\mathcal{D} \label{eq:ra_c6}
    \end{align}
\end{subequations}
where $P_{i}^{c}$ and $P_{j}^{d}$ denote the transmit power of CU $i$ and that of IBR-D2D $j$ respectively, and $\epsilon_{i}^{c}$ and $\epsilon_{j}^{d}$ denote the SINR of CU $i$ and that of IBR-D2D $j$ respectively. $\rho_{i,j}$ is the resource reuse indicator for CU $i$ and IBR-D2D $j$ , and $\rho_{i,j}=1$ when IBR-D2D $j$ reuses the spectrum of CU $i$; otherwise, $\rho_{i,j}=0$. $\epsilon_{i,\min}^{c}$ and $\epsilon_{j,\min}^{d}$ denote the minimum SINR requirements of CU $i$ and IBR-D2D $j$, respectively, and $P_{\max}^{c}$ and $P_{\max}^{d}$ are the maximum transit power of CUs and IBR-D2Ds. $g_{i,B},g_{j}$ are channel gains of non-D2D links and D2D links, while $h_{i,j},h_{j,B}$ are channel gains of interference links. To simplify the expression, we have $g_{i,B}=k_{i,B}L_{i,B}^{-\alpha}$, where $k_{i,B}$ denotes the synthetic gain including the fading and $L_{i,B}^{-\alpha}$ is a distance-dependent pathloss term \cite{feng2013device}. In this paper, we assume each IBR-D2D can find at least one reuse partner, i.e., $\mathcal{R}_j\neq\emptyset$ by define $\mathcal{R}_j$ as the set of IBR-D2D $j$'s reuse candidates.

The minimum SINR requirements $\epsilon_{i,\min}^{c}$ and $\epsilon_{j,\min}^{d}$ are necessary to satisfy the communication latency requirement for CPPS regulation and monitoring \cite{kong2020multicell}. Let $\tau$ be the maximum acceptable latency for a message, and all control traffic is shaped by a token bucket with rate $\phi$ bits per second and bucket size $\omega$ bits. Then, in order to satisfy the latency requirement $\epsilon_{i,\min}^{c}=\epsilon_{j,\min}^{d}=2^{\frac{\omega + \phi\tau}{W\tau}}-1$ with $W$ denoting the channel bandwidth.

In optimization \eqref{eq:ra_obj}, constraints \eqref{eq:ra_c1} and \eqref{eq:ra_c2} are the QoS requirements of CUs and IBR-D2Ds, respectively. Constraint \eqref{eq:ra_c3} ensures that the resource of an existing CU can be shared at most by one IBR-D2D, while constraint \eqref{eq:ra_c4} limits that a IBR-D2D shares at most one existing CU's resource. Constraints \eqref{eq:ra_c3} and \eqref{eq:ra_c4} ensure the exclusive resource reuse and reduce the interference complexity brought by involving D2D communications. Constraints \eqref{eq:ra_c5} and \eqref{eq:ra_c6} ensure the transmit powers of CUs and IBR-D2Ds are within the maximum limits.
It should be noted that the optimization \eqref{eq:ra_obj} is a nonlinear optimization problem, hence in the following it will be divided into three steps.

\begin{figure}[!t]
\centering
\includegraphics[width=0.8\columnwidth]{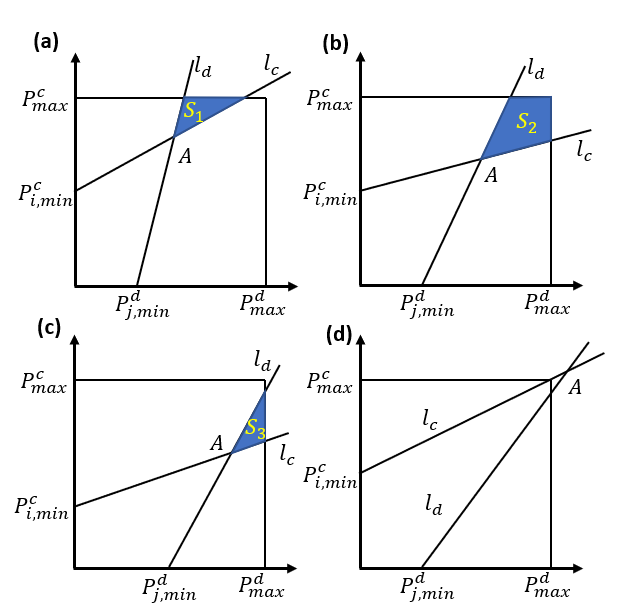}
\caption{Feasible reuse operation areas between one active CU and one IBR-D2D.}
\label{fig:feasible area}
\end{figure}

In the first step, each IBR-D2D finds its reuse partner to determine whether there exists at least one reuse candidate or not. If IBR-D2D $j$ can share spectrum with CU $i$, those constraints \eqref{eq:ra_c1}, \eqref{eq:ra_c2}, \eqref{eq:ra_c5} and \eqref{eq:ra_c6} must be satisfied. These constraints can be expressed in a geometric form by setting $P_{i}^{c},P_{j}^{d}$ ($y$ axis, $x$ axis respectively) as coordinate variables in a two dimensional axis. As shown in \cref{fig:feasible area}, feasible resource allocation areas for D2D reuse candidates are detailed, and intersection points with axes are $(0,P_{i,\min}^{c}),(P_{j,\min}^{d},0)$ with
\begin{align*}
    P_{i,\min}^{c}=\frac{\epsilon_{i,\min}^{c}\sigma^{2}}{g_{i,B}}, \quad P_{j,\min}^{d}=\frac{\epsilon_{j,\min}^{d}\sigma^{2}}{g_{j}}
\end{align*}
The area on the right of line $l_d$ and that above line $l_c$ are where respectively the minimum SINRs of IBR-D2D $j$ and CU $i$ are satisfied. Hence, from \cref{fig:feasible area}, we can find point A is located inside the first quadrant only when the slope of $l_d$ is larger than that of $l_c$, i.e., $\frac{\epsilon_{i,\min}^{c}h_{j,B}}{g_{i,B}}<\frac{g_{j}}{\epsilon_{j,\min}^{d}h_{i,j}}$, which is necessary condition for CU $i$ and IBR-D2D $j$. Then we define the intersection point A $(P_{j,A}^{d},P_{i,A}^{c})$, which can be found by
\begin{align*}
    \frac{P_{i,A}^{c}g_{i,B}}{\sigma^{2}+P_{j,A}^{d}h_{i,B}}=\epsilon_{i,\min}^{c},\quad
    \frac{P_{j,A}^{d}g_{j}}{\sigma^{2}+P_{i,A}^{c}h_{i,j}}=\epsilon_{j,\min}^{d}
\end{align*}
i.e.,
\begin{align}
    \left\{\begin{aligned}
    P_{i,A}^{c}=\frac{(g_{j}\epsilon_{i,\min}^{c}+h_{j,B}\epsilon_{i,\min}^{c}\epsilon_{j,\min}^{d})\sigma^{2}}{g_{j}g_{i,B}-\epsilon_{i,\min}^{c}\epsilon_{j,\min}^{d}h_{i,j}h_{j,B}} \\
    P_{j,A}^{d}=\frac{(g_{i,B}\epsilon_{j,\min}^{d}+h_{i,j}\epsilon_{i,\min}^{c}\epsilon_{j,\min}^{d})\sigma^{2}}{g_{j}g_{i,B}-\epsilon_{i,\min}^{c}\epsilon_{j,\min}^{d}h_{i,j}h_{j,B}}
    \end{aligned}\right.\label{eq:pointA}
\end{align}
Then, only if the point A (intersection point of lines $l_c,l_d$) is inside the square constrained by $P_{\max}^{c},P_{\max}^{d}$, CU $i$ is a reuse partner of IBR-D2D $j$, i.e., $0<P_{i,A}^{c}\leq P_{\max}^{c},0<P_{j,A}^{d}\leq P_{\max}^{d}$.

After the first step, we can obtain  $\mathcal{R}_{j}$ for each IBR-D2D $j$. The next step is to find the optimal power allocation for each reuse partner, i.e., solving $(P_{i}^{c*},P_{j}^{d*})=\argmax\;f(P_{i}^{c},P_{j}^{d})=\left[\log_2(1+\epsilon_i^c)+\log_2(1+\epsilon_j^d)\right]$ based on the feasible areas in \cref{fig:feasible area}. To explicitly express the optimal power allocation, we have the following piece-wise format
\begin{align}\begin{aligned}
	&(P_{i}^{c*},P_{j}^{d*})=\\ &\left\{\begin{aligned}
		&\argmax_{(P_{i}^{c},P_{j}^{d})\in \mathcal{S}_1}\;f(P_{i}^{c},P_{j}^{d}),\;& \left\{\begin{aligned}
        \frac{P_{\max}^{c}g_{i,B}}{\sigma^2+P_{\max}^{d}h_{j,B}}<\epsilon_{i,\min}^{c} \\ \frac{P_{\max}^{d}g_{j}}{\sigma^2+P_{\max}^{c}h_{i,j}}>\epsilon_{j,\min}^{d}
        \end{aligned}\right.\\
        &\argmax_{(P_{i}^{c},P_{j}^{d})\in \mathcal{S}_3}\;f(P_{i}^{c},P_{j}^{d}),\;& \left\{\begin{aligned}
        \frac{P_{\max}^{c}g_{i,B}}{\sigma^2+P_{\max}^{d}h_{j,B}}>\epsilon_{i,\min}^{c} \\ \frac{P_{\max}^{d}g_{j}}{\sigma^2+P_{\max}^{c}h_{i,j}}<\epsilon_{j,\min}^{d}
        \end{aligned}\right.\\
        &\argmax_{(P_{i}^{c},P_{j}^{d})\in \mathcal{S}_2}\;f(P_{i}^{c},P_{j}^{d}),\;& \left\{\begin{aligned}
        \frac{P_{\max}^{c}g_{i,B}}{\sigma^2+P_{\max}^{d}h_{j,B}}\geq\epsilon_{i,\min}^{c} \\ \frac{P_{\max}^{d}g_{j}}{\sigma^2+P_{\max}^{c}h_{i,j}}\geq\epsilon_{j,\min}^{d}
        \end{aligned}\right.
	\end{aligned}\right.
\end{aligned}\label{eq:power_opt}\end{align}
where $\mathcal{S}_1,\mathcal{S}_2,\mathcal{S}_3$ denote the feasible areas (a), (b), (c) respectively in \cref{fig:feasible area}. It should be noted that although \eqref{eq:power_opt} is optimized in areas, but the optimal powers for CU $i$ and D2D $j$ are located on corners of each area with at least one side at maximum power \cite{feng2013device}. Hence, it is efficient to solve \eqref{eq:power_opt} only via comparing $f(P_{i}^{c},P_{j}^{d})$ values of two or three points depending on piece-wise cases.

After obtaining reuse candidate sets $\mathcal{R}_{j}$ with the optimal power allocation, we need to match IBR-D2Ds with CUs to ensure that each IBR-D2D can find one reuse partner. Different IBR-D2Ds may have a same reuse partner, thus we need to quantify the sum rate enhancement to guide the best matching solution. Hence, for each IBR-D2D $j$, an ergodic calculation of sum data rate increment (of each CU in the set $\mathcal{R}_{j}$) is defined by $\Delta_{i,j}$
\begin{align}
	\Delta_{i,j}=\left\{\begin{aligned}
		&f(P_{i}^{c*},P_{j}^{d*})-\log_2(1+\frac{P_{\max}^{c}g_{i,B}}{\sigma^2}),\;&i\in\mathcal{R}_{j}\\
        &0,\;&i\notin\mathcal{R}_{j}
	\end{aligned}\right.
\end{align}
Thus, the IBR-D2D matching problem can be formulated as
\begin{subequations}
    \begin{align}
        & \max_{\rho_{i,j}}\quad \sum_{i\in\mathcal{C},j\in\mathcal{D}}{\rho_{i,j}\Delta_{i,j}} \tag {6} \label{eq:match_obj}\\
        & \mathrm{subject\ to:} \notag \\
        & \qquad\sum_{j}{\rho_{i,j}}\leq1,\rho_{i,j}\in\{0,1\},\forall\;i\in\mathcal{C} \label{eq:match_c1}\\
        & \qquad\sum_{i}{\rho_{i,j}}=1,\rho_{i,j}\in\{0,1\},\forall\;j\in\mathcal{D} \label{eq:match_c2}
    \end{align}
\end{subequations}
which turns to be a maximum weight bipartite matching problem and can be solved by the Hungarian algorithm \cite{west2001introduction} efficiently.

In each region, the above three steps are applied to gather IBRs state information via D2D-assisted wireless communications. The wireless resources managed by optimization problem \eqref{eq:ra_obj} (solved step-by-step using \eqref{eq:pointA}--\eqref{eq:match_obj}) can be allocated separately in terms of multiple regions.

\section{Simulation Results\label{sec:4}}
In this section, the wireless communication that supports renewable IBR integration is elaborated. We start a wireless communication solution for one cell network, where IBR-CUs and IBR-D2Ds are located within radius $r$. In the simulation, the total bandwidth is shared by IBR-CUs, while IBR-D2Ds reuse a part of bandwidth. If the total bandwidth is reused by IBR-D2Ds, IBR-CUs and IBR-D2Ds have the equal number, as an extreme case of D2D communication reuse scenario. The main simulation parameters are detailed in \cref{table:parameters_ra} partially extracted from~\cite{kong2020multicell,liang2017resource}, and the channel models are based on the urban scenario of WINNER II Channel Models \cite{bultitude20074} and further concluded as TABLE IV in \cite{liang2017resource}.
\\\vspace{-2em}
\begin{table}[!htb]
	\renewcommand{\arraystretch}{1.3}
	\centering
	\caption{Simulation Parameters}\vspace{-0.8em}
	\resizebox{0.48\textwidth}{!}{
	\begin{tabular}{l|l||l|l}
		\hline\hline
		Parameter & Value & Parameter & Value \\
		\hline
		Carrier frequency & 2 GHz & Maximum IBR-CU & \multirow{2}{*}{23 dBm} \\
		Bandwidth & 4 MHz & transmit power $P_{\max}^c$ & \\
		Cell radius $r$ & 500 m & Maximum IBR-D2D & \multirow{2}{*}{23 dBm} \\
		BS antenna height & 25 m & transmit power $P_{\max}^d$ &\\
		BS antenna gain & 8 dBi & Noise power $\sigma^2$ & -114 dBm \\
		BS receiver noise figure & 5 dB & Packet size & 32 bytes \\
		IBR antenna height & 1.5 m & Token bucket size $\omega$ & 60 packets \\
		IBR antenna gain & 3 dBi & Token bucket rate $\phi$ & 60 packets/s \\
		IBR receiver noise figure & 9 dB & Latency requirement $\tau$ & 20 ms \\
		\hline\hline
	\end{tabular}
	}\label{table:parameters_ra}
\end{table}

\begin{figure}[!htb]
\centering
\includegraphics[width=0.8\columnwidth]{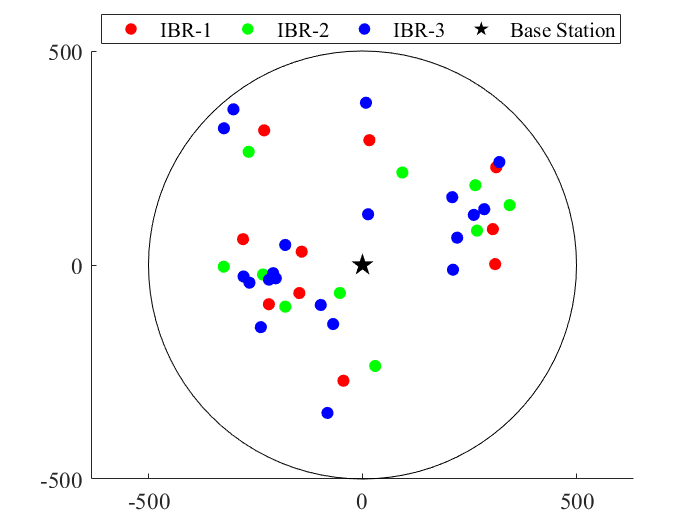}
\caption{Geographical locations of base station, IBRs in a cellular network.}
\label{fig:0}
\end{figure}

\begin{figure*}[!htb]
\centering
\includegraphics[width=1.6\columnwidth]{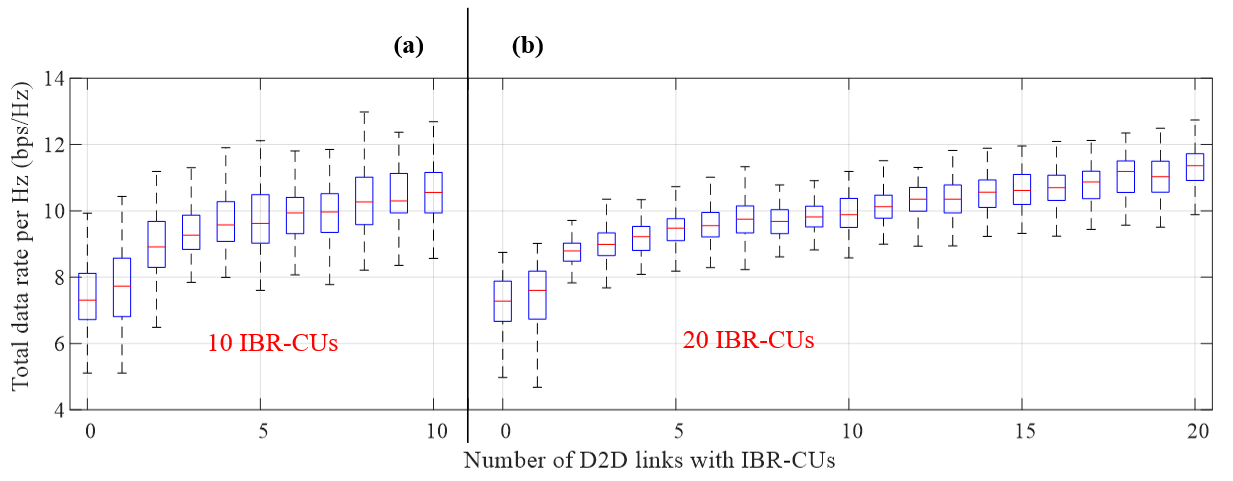}
\caption{Total throughput capacity with an increasing number of IBR-D2Ds: 10 IBR-CUs case on the left, 20 IBR-CUs case on the right.}
\label{fig:1}
\end{figure*}

To evaluate the wireless communication throughput capacity including D2D communications, we analyse the performance in terms of bandwidth utilization rate and accessible IBRs using D2D communications. We start from a case including 10 IBR-CUs, which can accommodate at most 10 IBR-D2Ds, and the locations of IBR-CUs and IBR-D2Ds are randomly located in the cellular network coverage. A typical location are depicted in red dots and green dots respectively of \cref{fig:0}, and we generate a random distribution of 100 different cases. The unit data rates (bps/Hz) of the cellular network supporting IBRs information flow are outlined in \cref{fig:1}(a). As the number of IBR-D2Ds (the number of connected IBR-D2Ds, green dots in \cref{fig:0}) increases, the total data rate per Hz increases remarkably, leading to an increasing bandwidth utilization. This demonstrates D2D communications can improve the performance of wireless communication and further accommodate more renewable IBRs. Owing to the fact that the bandwidth can be dynamical allocated to form a time-varying communication network, the capability of supporting plug-and-play operation of mobile IBRs can be provided from both sides of data rate and network topology.

Then, the benefits of D2D communication is further demonstrated inspired by the bandwidth slicing \cite{zhou2018bandwidth}. We set IBRs denoted by both red and green dots in \cref{fig:0} as IBR-CUs with half channel bandwidth. It should be emphaisized that if 20 IBRs are set as CUs without reuse partners, the unit data rate drops dramatically (see the first column of \cref{fig:1}(b)). Similarly, D2D communications improve the wireless system performance with an increasing number of IBR-D2Ds, the locations of which are depicted as blue dots in \cref{fig:0}. Thanks to the bandwidth slicing technology, a cellular network has the capability of accommodating both legacy and new IBRs by appropriate bandwidth allocation reconfiguration from the base station side.

\section{Concluding Remark\label{sec:5}}
This paper proposes a hierarchical structure that combines the advantages of both centralised control and distributed control via the advanced wireless communication system to facilitate a cost-effective information scheme supporting IBRs integration. The regional controller, assisted by base station, enables the coordination of legacy and new IBRs. The base station dynamically allocate wireless bandwidth to enable a time-varying communication network, allowing plug-and-play operations from the angle of cyberspace. The D2D communications improve the bandwidth efficiency by reusing channels to accommodate extra IBRs. On the other side, in addition to that the EMS or SCADA can offline dispatch regional controllers in a traditional centralised and long-time-scale manner, regional controllers can communicate with each other via distributed feedback control, which will enhance the controllability of IBRs in a short time scale. In the end, the wireless system performance supporting regional IBR aggregations is evaluated. Through the well developed control algorithms, the CPPS can be controlled efficiently and effectively.Finally, the findings from this work provide an important step toward enabling energy access for global nomadic communities based on the improved compatibility of the proposed framework with a migratory lifestyle, as opposed to existing solutions. The long-term planning of such a highly complex and uncertain system, however, will necessitate the development of novel data-driven approaches focused on flexibility, for instance as shown by \cite{caputo2022analyzing} in a different energy system context.



\bibliographystyle{IEEEtran}
\bibliography{ref}

\end{document}